\documentclass[conference]{IEEEtran}
\ifCLASSINFOpdf
\else
\fi
\usepackage{algorithm}
\usepackage{algorithm,algpseudocode}
\usepackage{graphicx}
\usepackage{array}
\usepackage{float}
\usepackage{lipsum}
\usepackage{authblk}
\usepackage{authblk}% The option is for block layout
\usepackage{hyperref}% For email address

\begin{document}
\title{Limitations of OpenFlow Topology Discovery Protocol}
% 
% \author{Abdelhadi Azzouni}
% \author{Jessica}
% \affil{Department 1 \\% If the blocks option of authblk is removed \\ is treated as ,
% School 1 \\
% \url{email1}}
% 
% \author{John}
% \author{Chris}
% \affil{Department 2 \\
% School 2 \\
% \url{email2}}
% 
% \author{Anna}
% \affil{Department 3 \\
% School 3 \\
% \url{email3}}

\author[1]{Abdelhadi Azzouni\thanks{abdelhadi.azzouni@lip6.fr}}
% \author[2]{Othmen Braham\thanks{othmen.braham@virtuor.fr}}
\author[1]{Nguyen Thi Mai Trang\thanks{thi-mai-trang.Nguyen@lip6.fr}}

\author[2]{Raouf Boutaba\thanks{E.E@university.edu}}
\author[1]{Guy Pujolle\thanks{guy.pujolle@lip6.fr}}

\affil[1]{LIP6 / UPMC; Paris, France  \{abdelhadi.azzouni,thi-mai-trang.nguyen,guy.pujolle\}@lip6.fr}
% \affil[2]{Virtuor; Paris, France  othmen.braham@virtuor.fr}
\affil[2]{University of Waterloo; Waterloo, ON, Canada  rboutaba@uwaterloo.ca }
\maketitle

\begin{abstract}

OpenFlow Discovery Protocol (OFDP) is the de-facto protocol used by OpenFlow controllers to discover the underlying topology.
In this paper, we show that OFDP has some serious security, efficiency and functionality limitations that make it non suitable for production deployments.
Instead, we briefly introduce sOFTD, a new discovery protocol with a built-in security characteristics and which is more efficient than traditional OFDP.

%\boldmath

\end{abstract}
% This preserves the distinction between vect	ors and scalars. However,
% \begin{keywords}
% Software-Defined Networking, OpenFlow, Control Plane, security.
%  
% \end{keywords}
% 

 {\bf { \it keywords - }}
Software-Defined Networking, OpenFlow, Topology Discovery, security.

% \fi
\IEEEpeerreviewmaketitle

\section{Introduction}


The separation between the control plane and the data plane introduced by Software-Defined Networking (SDN) allows 
operators to employ quite damn, remarkably cheap but very fast hardware to forward packets, moving the control logic to a centralized and much smarter entity
called controller. The controller plays the role of an operating system of the network. It abstracts the 
underlying forwarding hardware details and offers high level APIs that the network admins leverage to program their networks.
% The controller is responsible for compiling high level network policies expressed by administrators to 
% elementary instructions that the forwarding hardware can understand and execute. 
One of the fundamental functions that a controller must offer is an accurate, near real time visibility of the network topology. This function is known as Topology Discovery. 
Topology discovery in SDN is more sensitive compared to traditional networks based on Link-State routing protocols like OSPF.
In SDN, To discover the network topology, all current OpenFlow controllers implement the same protocol OFDP (OpenFlow Discovery Protocol).

\begin{figure} [h]
\centering
   \includegraphics[scale=0.4]{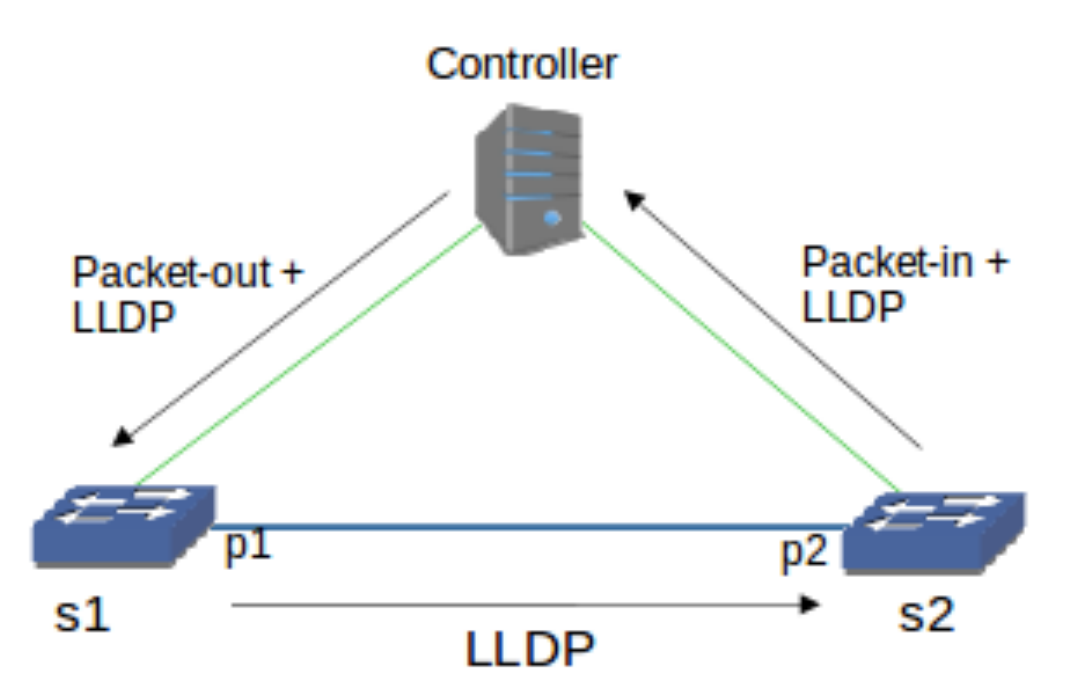}
%    \vspace{-1em}
   \caption{\label{ofdp} Discovering a unidirectional link in OFDP}
%    \vspace{-1em}
\end{figure}

Figure \ref{ofdp} shows how OFDP works; To discover the unidirectional link s1 $\rightarrow$  s2, 
the controller encapsulates a LLDP packet in a Packet-out message and sends it to s1. The packet-out contains instruction for s1 to send the LLDP packet to s2 via port p1.
By receiving the LLDP packet via port p2, s2 encapsulates it in a Packet-in message and sends it back to the controller. 
The controller receives the LLDP packet and concludes that there is a unidirectional link from s1 to s2.
The same process is performed to discover the opposite direction s2 $\rightarrow$  s1 as well as all other links in the network.
Note that, OFDP packets are sent to a "normal" multicast MAC (01:23:00:00:00:01) to avoid being swallowed by 802.1d compliant switches.

In dynamic networks like large data-centers and multi-tenant cloud networks, keeping an up-to-date visibility of the topology is a critical function; 
Switches leave and join the network dynamically creating changes in the topology which affects routing decisions that the controller has to make continuously.
To remain up-to-date, the controller needs to repeat the process described in figure \ref{ofdp} periodically. 
The period separating two discovery rounds must be chosen carefully based on the dynamicity, size and capacity of the network;
A 10 seconds period might not be suitable for a highly dynamic network as it may introduce a delay of up to 10 seconds.
A short period (e.g. 3 seconds) also might not be suitable for a less-dynamic large size network as the large number 
of frequent discovery packets may exhaust controller's resources. 
Put together, every discovery-round period $T$, the controller sends $\sum_{i=1}^{n} p_i$ (where $n$ is 
the number of switches and $p_i$ is the number of ports in switch $i$) Packet-out messages and receives $2L$ Packet-in messages.
\cite{simplepaper} proposes to reduce the number of Packet-out messages to $n$ by rewriting LLDP packet-headers in the switch.

A non optimized or buggy topology discovery mechanism can affect routing logic and drastically reduce network performance.
Our main goal in this paper is to demonstrate that $OFDP$ has serious, non-solved yet, security and performance problems, 
then we briefly introduce $sOFTD$  (secure and efficient OpenFlow Topology Discovery), 
a secure alternative that is more efficient than $OFDP$.

The remainder of this paper is organized as follows: In section \ref{whyofdpbad} we demonstrate why $OFDP$ shouldn't be implemented in production networks.
We introduce our alternative protocol $sOFTD$ in section \ref{introducingsoftd} and we conclude the paper in section \ref{conclusion}

% in the form of a graph, showing which nodes are connected to which other nodes. 
% Each node then independently calculates the next best logical path from it to every possible destination in the network. 
% The collection of best paths will then form the node's routing table.

\section{Why OFDP shouldn't be implemented in production networks} \label{whyofdpbad} 
\subsection{OFDP is not secure}
As implemented by all controllers we have tested (OpenDaylight \cite{odl}, Floodlight \cite{floodl2}, NOX \cite{nox}, POX \cite{nox}, 
Beacon \cite{beacon}, Ryu \cite{ryu} and Cisco Open SDN Controller \cite{opensdn}), 
OFDP uses clear, non authenticated LLDP packets to detect links between switches which makes it vulnerable to a number of attacks: \\

\begin{figure}[h] \label{lldppacket}
\centering
   \includegraphics[scale=0.25]{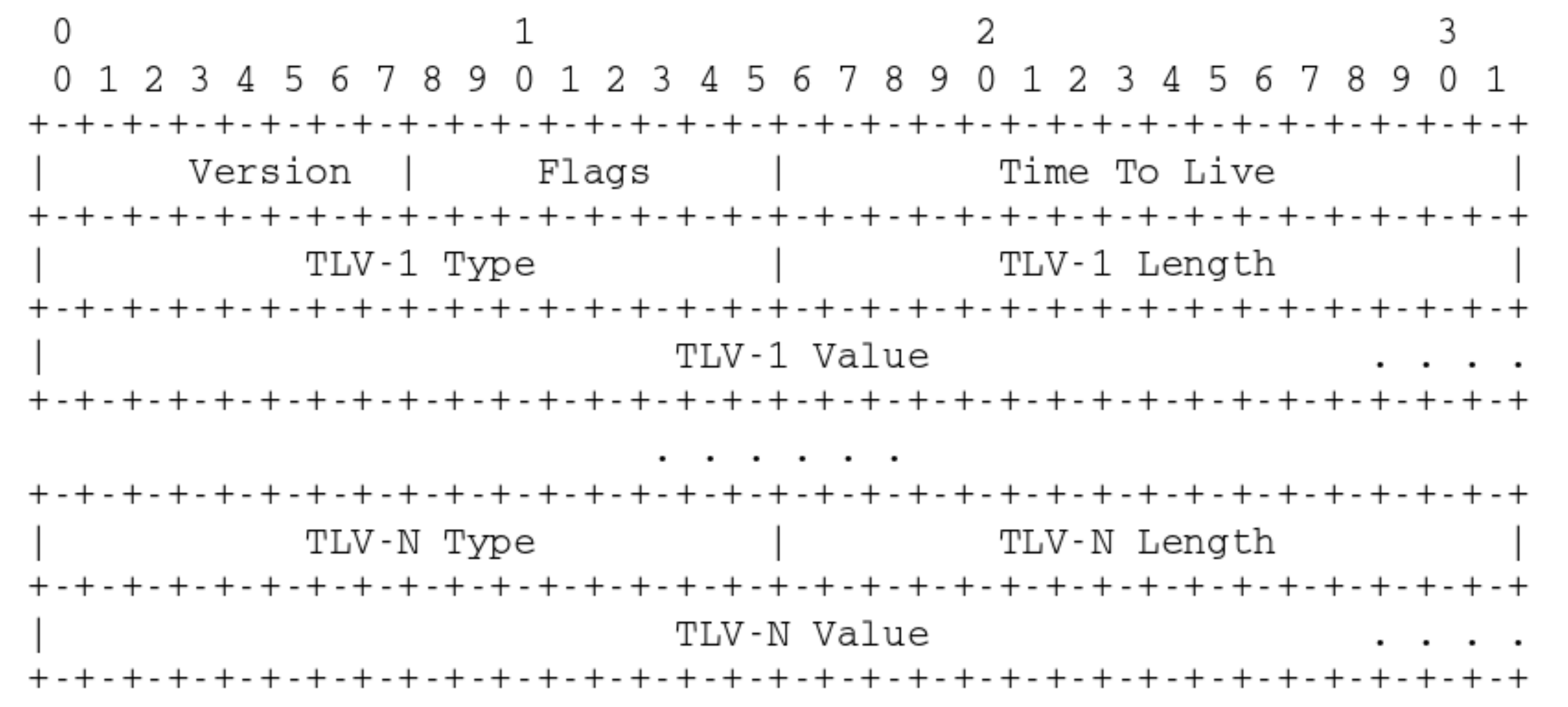}
   \vspace{-1em}
   \caption{\label{fig:lldppacket} LLDP packet format \cite{rfclldp}}
%    \vspace{-1em}
\end{figure}

\textbf{Switch spoofing.} As described in figure \ref{fig:lldppacket}, each LLDP packet contains a version field, flags, TTL and TLVs (Type-length-value) for information advertisement. 
Mandatory TLVs in OFDP are $Chssis Subtype$ and $Port Subtype$ to track packets. In figure \ref{ofdp}, 
the LLDP packet sent by switch s2 to the controller contains the tuple ($chassis Subtype=switch1 ID, Port Subtype=p1$), hence the controller 
will detect that this is the 
same packet he sent to switch s1 with p1 as out-port. 

% \vspace{+0em}

\begin{figure}[h] \label{lldppacket}
\centering
   \includegraphics[scale=0.3]{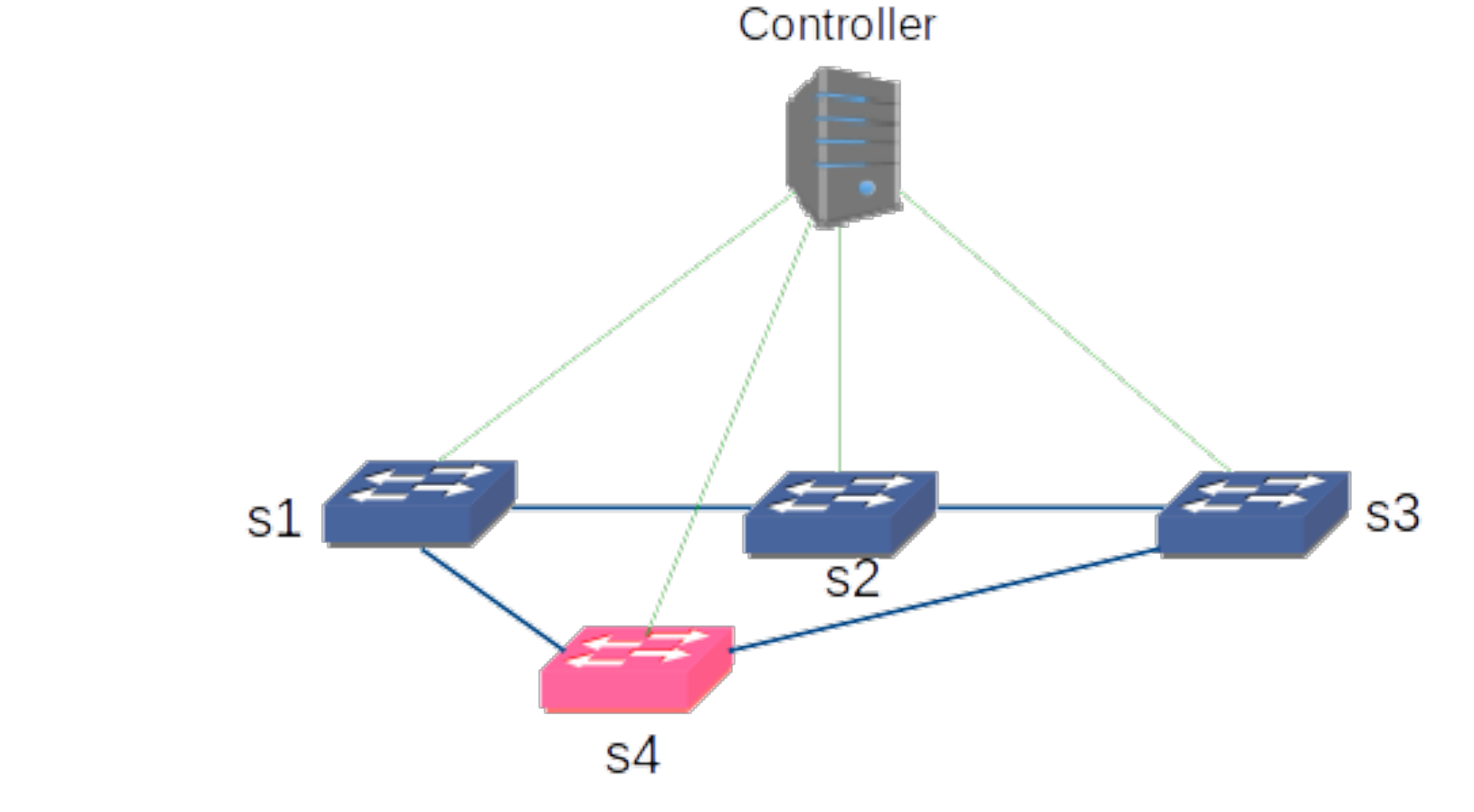}
   \vspace{-1em}
   \caption{\label{fig:switchspoof} Switch spoofing attack}
%    \vspace{-1em}
\end{figure} 
% \vspace{+0em}

\begin{figure} \label{wiresharkpox} 
\centering
   \includegraphics[scale=0.4]{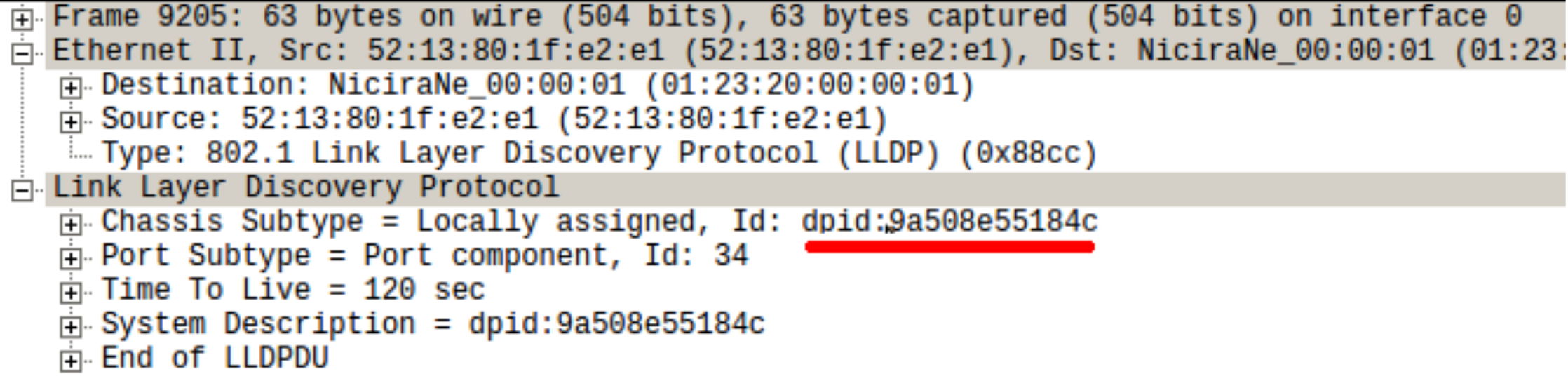}
%    \vspace{-1em}
   \caption{\label{fig:wiresharkpox} LLDP content used by POX controller}
%    \vspace{-1em}
\end{figure}

The problem is that all controllers we have tested 
set $chassis Subtype$ value to the MAC address of the local port of the switch (figure \ref{fig:wiresharkpox}), which makes it easy for an adversary to spoof that switch since 
controllers use that MAC address as a unique identifier of the switch. By intercepting clear LLDP packets containing MAC addresses, 
a malicious switch can spoof other switches to falsify the controller's topology graph. 
In the example shown in figure \ref{fig:switchspoof}, s4 intercepts LLDP packets from s1 containing s1's local port MAC address. Now, s4
can use it as its own MAC and reconnect to the controller as s1 messing up the controller's topology graph (e.g. the controller adds nonexistent links $s1 \rightarrow s3$ and $s3 \rightarrow s1$).
We have tested the switch spoofing attack successfully against Opendaylight and Floodlight.  \\

% \vspace{+0em}

\begin{figure} \label{linkfabric} 
\centering
   \includegraphics[scale=0.2]{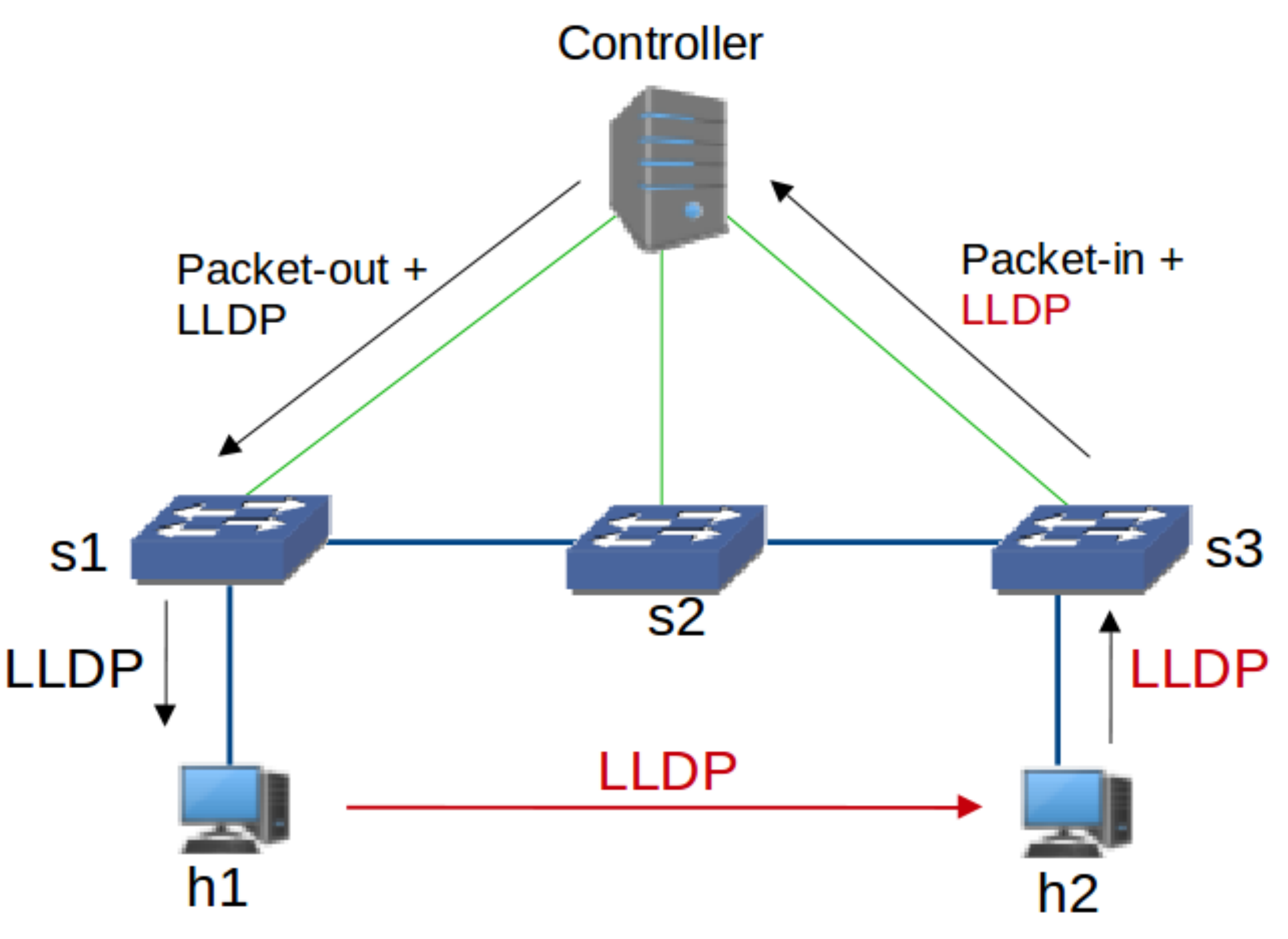}
   \caption{\label{fig:linkfabric} Link Fabrication attack}
%    \vspace{-1em}
\end{figure}

\textbf{Link Fabrication.} \cite{poisonning} and \cite{insecurity} pointed out that OFDP is vulnerable to link fabrication attacks;
In figure \ref{fig:linkfabric}, the adversary has control over two end-hosts h1 and h2 connected to switches s1 and s3. 
h1 sends the LLDP packets received from s1 to h2 through an out-of-band connection (could be a tunnel over s2 for example), and h2 replicates them to s3.
The controller receives the LLDP packets from s3 and creates a link between s1 and s3. While not detected, 
the fake link pushes the controller into wrong routing decisions that affects all communications involving s1 and s3. 
If the attacker has control only over h1, but knows the DPID of s3 then he still can fabricate a unidirectional link s3 $\rightarrow$ s1 by injecting 
fake LLDP packets into s1.

% \vspace

Another form of link fabrication is by LLDP injection; By monitoring the traffic, the adversary gets the LLDP content used by the controller. 
Then, he/she injects the same LLDP packets into the network creating bogus links between switches or between the adversary hosts and switches.

\cite{poisonning} proposes to authenticate the LLDP packets by adding a key-Hash Message Authentication Code (HMAC) as an optional TLV in LLDP packets.
As mentioned by the authors, this technique only works against fake LLDP injection but not against link fabrication by packet duplication (Figure \ref{fig:linkfabric}). 
\cite{insecurity} proposes a similar technique but using dynamic keys: a unique key for each LLDP packet assuming that h2, for example, uses only one example of 
LLDP packets received from h1 to generate future fake packets. However, an attacker controlling both hosts can permanently forward captured LLDP packets from h1 to h2 and from h2 to h1
and inject them back into switches without any modification.

\begin{figure} \label{switchspoof} 
\centering
   \includegraphics[scale=0.3]{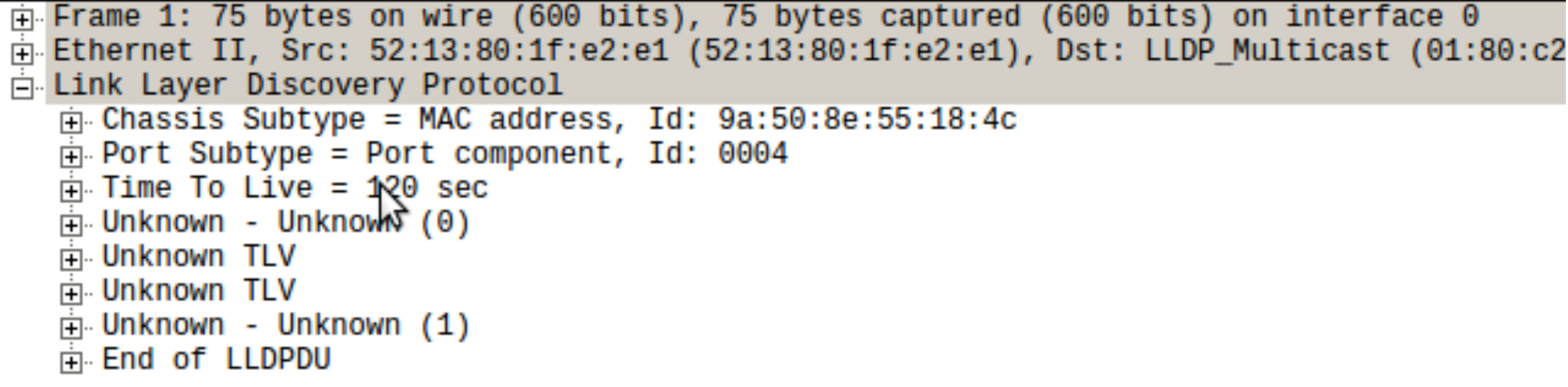}
%    \vspace{-1em}
   \caption{\label{fig:wiresharkflood} LLDP content used by Floodlight controller}
%    \vspace{-1em}
\end{figure}

\textbf{Controller fingerprinting.} As we explained in a previous work \cite{sdnfingerprint}, the LLDP content is different from one controller to another which
allows fingerprinting attacks on SDN controllers. An adversary (h1 in figure \ref{fig:linkfabric}) matches the LLDP content he receives 
from s1 (LLDP packets originate from the controller) against a controller signature database to detect which controller is managing the network. Such information is 
very useful to launch specific and more efficient attacks on the controller. Figures \ref{fig:wiresharkpox} and \ref{fig:wiresharkflood} present the LLDP content of controllers POX and 
Floodlight respectively. 
Note also that the controllers use different default discovery-round periods which offers another way to differentiate between controllers. 
Although it is possible that network admins change both discovery-round period and LLDP content, it is more likely that the default values are kept unmodified.

% \vspace

\begin{figure} \label{switchspoof} 
\centering
   \includegraphics[scale=0.4]{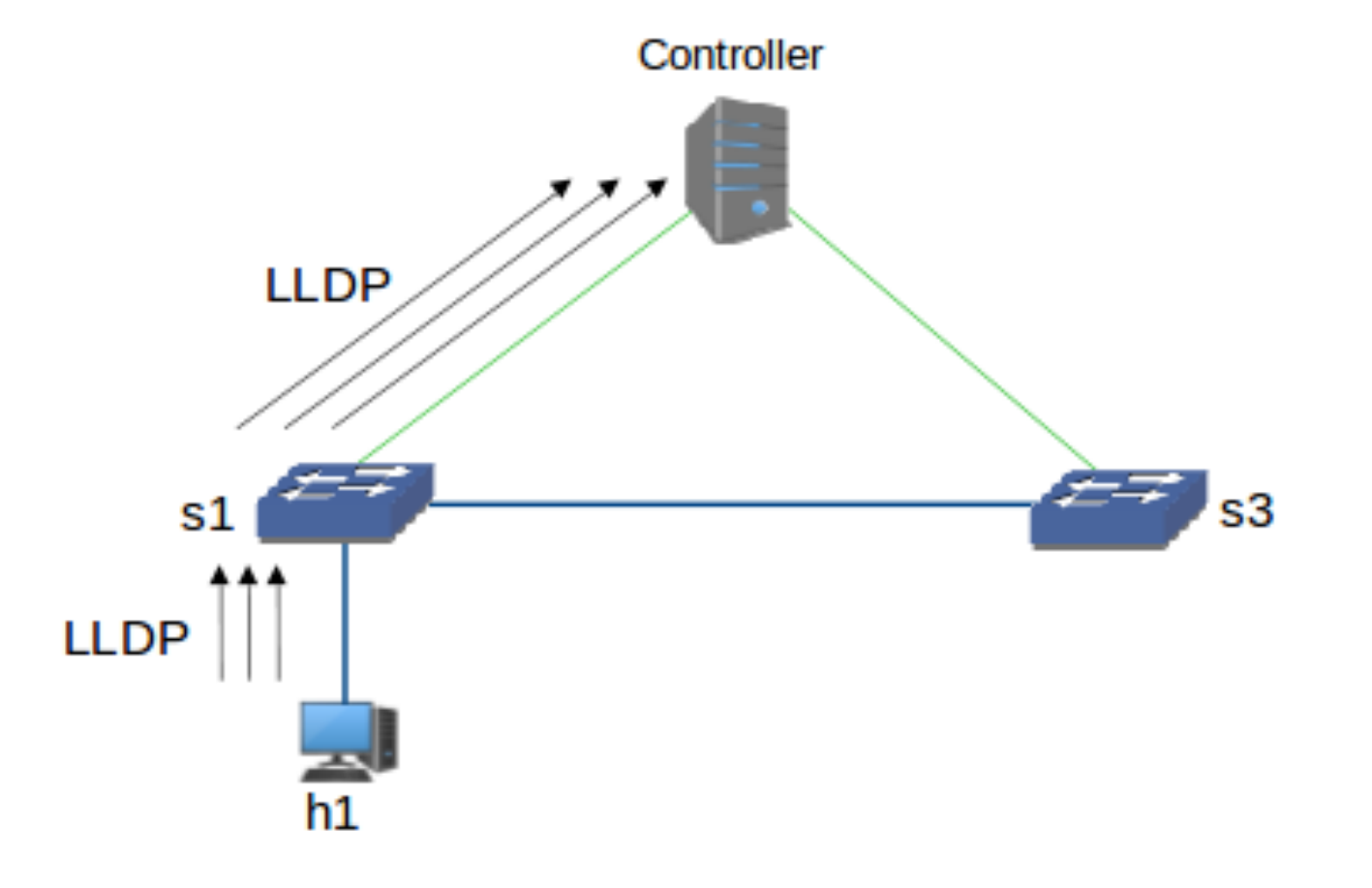}
%    \vspace{-1em}
   \caption{\label{fig:lldpfloodattack} LLDP flood attack}
%    \vspace{-1em}
\end{figure}

\textbf{LLDP Flood.} This is a form of DoS attack where an adversary generates enough fake LLDP packets to exhaust the controller resources.
In figure \ref{fig:lldpfloodattack}, host h1 generates large number of LLDP packets and send them to s1 which has a rule to forward 
every LLDP packet to the controller. Hence, a large number of LLDP packets can exhaust the link connecting the switch to the controller 
as well as the controller resources. Basic countermeasure methods like port blocking or packet filtering may not be effective, especially in the
case of very dynamic environments (e.g. multi-tenant cloud) since connected hosts and switches change frequently, which may result in preventing legitimate LLDP packets from reaching the 
controller.

\subsection{OFDP is not efficient}
By using OFDP, the controller periodically sends many packets to every
switch in the network, which could result in performance decrease of the data
plane. Experiments made on different controllers \cite{necwhitepaper} show that 
when the network size (i.e. number of switches) exceeds some threshold, 
running the discovery
module alone results in significant increase of the controller's CPU usage and considerable decrease
in network performance.

\subsection{Other issues}

Other issues with OFDP include that it may not 
reliably work for heavily loaded links because discovery packets might get dropped or delayed. 
Moreover, when using OFDP in a multi-controller SDN network (e.g. running several guest controllers through FlowVisor), 
discovery cost increases linearly as more controllers are added.

\section{Introducing sOFTD: Secure OpenFlow Topology Discovery}\label{introducingsoftd}
The main idea behind sOFTD is to move a part of the discovery process from the
controller to the switch by introducing minimal changes to the OpenFlow switch design. The main design characteristics of sOFTD are as follows: 

\begin{itemize}
 \item sOFTD adds a port-liveness-detection mechanism (BFD) \cite{bfd} to the switch
 \item sOFTD uses OpenFlow FAST-FAILOVER groups (optional in OpenFlow 1.1+) to watch switch ports for connection updates
 \item sOFTD enables the switch to inform the controller about port connectivity updates
 \item The switch has a rule ("drop lldp") to drop every LLDP packet to
prevent LLDP flood attack
 \item The controller sends encrypted LLDP packets only when a switch-port
connectivity update occurs and the LLDP packets are sent only to the concerned
switches
 \item LLDP packets are preceded by rules (with hard timeout = 1s) to
forward them back to the controller. These rules have a priority
higher than "drop lldp" rules
\end{itemize}

Since sOFTD do not send periodic discovery packets, preliminary resuts show significantly better performance than OFDP. 
We keep implementation details and results for a future work.

\section{Conclusion}\label{conclusion}
In this short paper, we demonstrated some serious security and efficiency problems in OpenFlow Discovery Protocol. 
Most of these problems are yet unsolved. As an alternative to OFDP, 
we briefly introduced our ongoing work, $sOFTD$ protocol, which consists of 
moving part of the discovery intelligence from the controller to the switch by introducing minimal changes to the OpenFlow switch design.


\begin{thebibliography}{1}
\bibitem{poisonning}
Hong, Sungmin, et al. "Poisoning Network Visibility in Software-Defined Networks: New Attacks and Countermeasures." NDSS. 2015.

\bibitem{rfclldp}
Congdon, P. (2002). Link layer discovery protocol and MIB. V1. 0 May 20. 2002, http://www. IEEE802.

\bibitem{insecurity}
Alharbi, Talal, Marius Portmann, and Farzaneh Pakzad. "The (In) Security of Topology Discovery in Software Defined Networks.
" Local Computer Networks (LCN), 2015 IEEE 40th Conference on. IEEE, 2015.

\bibitem{simplepaper}
Pakzad, Farzaneh, et al. "Efficient topology discovery in software defined networks." Signal Processing and Communication Systems (ICSPCS), 2014 8th International Conference on. IEEE, 2014.

\bibitem{odl}
Linux Foundation. "OpenDaylight". https://www.opendaylight.org/.

\bibitem{floodl2}
Project Floodlight. https://floodlight.atlassian.net/wiki/display/floodlight-controller/Supported+Topologies


\bibitem{nox}
Gude, Natasha, et al. ”NOX: towards an operating system for networks.
” ACM SIGCOMM Computer Communication Review 38.3 (2008): 105-110.

\bibitem{beacon}
Erickson, David. "The beacon openflow controller." In Proceedings of the second ACM SIGCOMM workshop on Hot 
topics in software defined networking, pp. 13-18. ACM, 2013.


\bibitem{ryu}
Ryu. http://osrg.github.com/ryu/

\bibitem{opensdn}
Cisco. "Cisco Open SDN Controller". http://www.cisco.com/c/en/us/products/cloud-systems-management/open-sdn-controller/index.html

\bibitem{sdnfingerprint}
Azzouni, A et al. "Fingerprinting OpenFlow controllers: The first step to attack an SDN control plane". GLOBECOM. 2016. To appear.

\bibitem{necwhitepaper}
IXIA and NEC. "White paper: SDN Controller Testing, Part 1". https://www.necam.com/docs/?id=2709888a-ecfd-4157-8849-1d18144a6dda

\bibitem{bfd}
IETF, Bidirectional Forwarding Detection (BFD),
https://tools.ietf.org/html/rfc5880
% \bibitem{survey1}
% Ahmad, Ijaz, et al. "Security in software defined networks: a survey." Communications Surveys \& Tutorials, IEEE 17.4 (2015): 2317-2346.
% 
% \bibitem{survey2}
% Scott-Hayward, Sandra, Sriram Natarajan, and Sakir Sezer. "A survey of security in software defined networks." (2015).
% 
% \bibitem{ONF-sec}
% Open Networking Foundation. "SDN Security Considerations in the Data Center",
% Version 1.4.0 (Wire Protocol 0x05). October 14, 2013
% 
% \bibitem{nmap}
% NMAP. https://nmap.org/
% 
% \bibitem{zap}
% OWASP Zed Attack Proxy Project. https://www.owasp.org/index.php/OWASP\_Zed\_Attack\_Proxy\_Project
% 
% \bibitem{ONF}
% Open Networking Foundation. "Software-Defined Networking". https://www.opennetworking.org/sdn-resources/sdn-definition.
%   
% \bibitem{ONF-OF}
% Open Networking Foundation. "OpenFlow". https://www.opennetworking.org/sdn-resources/openflow.
% 
% \bibitem{OF-specif}
% Open Networking Foundation. "OpenFlow Switch Specification",
% Version 1.5.0 (Wire Protocol 0x06). December 19, 2014.
% 
% \bibitem{sdnscanner}
% S. Shi, G. Gun. “Attacking Software-Defined Networks: A First Feasibility Study”, ACM Proc. of HotSDN, Hong Kong, China, 2013.
%   
% \bibitem{inference}
% Leng, Junyuan, Yadong Zhou, Junjie Zhang, and Chengchen Hu. "An Inference Attack Model for Flow 
% Table Capacity and Usage: Exploiting the Vulnerability of Flow Table Overflow in Software-Defined Network." 
% arXiv preprint arXiv:1504.03095, 2015. 
% 
% \bibitem{floodl2}
% Project Floodlight. https://floodlight.atlassian.net/wiki/display/floodlight-
% controller/Supported+Topologies
% 
% \bibitem{odl}
% Linux Foundation. "OpenDaylight". https://www.opendaylight.org/.
% % 
% % \bibitem{rfcMIB}
% % Bierman, A., and K. Jones. "RFC 2922." Physical topology MIB, 2000.
% 
% \bibitem{nox}
% Gude, Natasha, et al. ”NOX: towards an operating system for networks.
% ” ACM SIGCOMM Computer Communication Review 38.3 (2008): 105-
% 110.
% 
% \bibitem{beacon}
% Erickson, David. "The beacon openflow controller." In Proceedings of the second ACM SIGCOMM workshop on Hot 
% topics in software defined networking, pp. 13-18. ACM, 2013.
% 
% \bibitem{beacon2}
% What is Beacon?. https://openflow.stanford.edu/display/Beacon/Home/
% 
% \bibitem{floodlight}
% Floodlight. http://Floodlight.openflowhub.org/
% 
% \bibitem{ryu}
% Ryu. http://osrg.github.com/ryu/


\end{thebibliography}
\end{document}